\documentclass[epsfig]{elsart}
\usepackage{graphicx,epsfig}
\usepackage{amsmath,amssymb}

\newcommand{\newc}{\newcommand}
\newc{\lra}{\leftrightarrow}
\newc{\beq}{\begin{equation}}
\newc{\eeq}{\end{equation}}
\newc{\barr}{\begin{eqnarray}}
\newc{\earr}{\end{eqnarray}}

\begin{document}
\date{\today}
\title {THE ROLE OF IONIZATION ELECTRONS IN DIRECT
NEUTRALINO DETECTION}
\author{ J.D. Vergados$^{1*}$ and H. Ejiri$^2$}
\address{
 1 Physics Department, University of Cyprus, 1678 Nicosia,
 Cyprus
$$and$$
 Theoretical Physics Division, T-6, LANL, P.O. Box 1663, Los Alamos, N.M. 87545,
 USA.}
 \footnote{Permanent address: University of Ioannina, Gr 451 10,
Ioannina, Greece.\\ E-mail:Vergados@cc.uoi.gr}
 \address{ 2 NS,  International Christian University, Osawa, Mitaka, Tokyo, 181-8585, Japan
 $$and$$
 Emeritus RCNP, Osaka University, Ibaraki,
 Osaka 567-0047,Japan}
\begin{frontmatter}
\begin{abstract}
In this paper we estimate the event rates of neutralino-nucleus
scattering leading to electron emission by the atomic ionization.
 We find that the branching ratio for detecting electrons vis a vis the
 traditional neutralino nucleus elastic scattering sensitively depends
 on the threshold energy. In the case of a light target and a neutralino mass of 100 GeV
  we estimate it to be around 10 percent assuming a reasonable threshold energy of 250 eV.
  \end{abstract}
\end{frontmatter}
\section{Introduction}
The combined MAXIMA-1 \cite{MAXIMA-1}, BOOMERANG \cite{BOOMERANG},
 DASI \cite{DASI} and COBE/DMR \cite{COBE} Cosmic Microwave Background (CMB)
observations as well as the recent WMAP data \cite{SPERGEL} imply
that the Universe is flat \cite{flat01} and that most of the
matter in the Universe is Dark, i.e. exotic. Crudely speaking  one
has:
$$\Omega_b=0.05, \Omega _{CDM}= 0.30, \Omega_{\Lambda}= 0.65$$
for the baryonic, dark matter and dark energy fractions
respectively.
 Furthermore, since
the non exotic component cannot exceed $40\%$ of the CDM ~\cite
{Benne}, there is room for the exotic WIMP (Weakly Interacting
Massive Particles).
  Many experiments are currently under way aiming at the direct detection
 of WIMP.  In fact the DAMA experiment ~\cite {BERNA2} has claimed the
  observation of  such events, which with better statistics
   have subsequently been interpreted
as a modulation signal \cite{BERNA1}. These data, however, are not
consistent with other recent experiments, see e.g. EDELWEISS
\cite{EDELWEISS} and CDMS \cite{CDMS}.

 Supersymmetry naturally provides candidates for the dark matter constituents
\cite{Jung},\cite{GOODWIT}-\cite{ELLROSZ}.
 In the most favored scenario of supersymmetry the
lightest supersymmetric particle (LSP) can be simply described as
a Majorana fermion, a linear combination of the neutral components
of the gauginos and higgsinos
\cite{Jung},\cite{GOODWIT}-\cite{WELLS}.

 The event rates, however, are expected to be quite low and the nuclear
recoil energies are extremely small. Thus one has to try to reduce
the background to the lowest possible level and understand
possible reaction induced backgrounds(for a brief discussion see
the appendix at the end of the paper). Furthermore one has to
search for characteristic signatures associated with this
reaction. Examples are the modulation of the event rates with the
motion of the Earth (modulation effect) and the correlation of the
observed rates of directionally sensitive experiments with the
motion of the sun \cite{SLSKA,DIREXP,JDV03}. Transitions to low energy excited
nuclear states have also been considered \cite{VQD03,EFO93}.

In the present paper we will explore another novel possibility,
namely detecting dark matter constituents by observing the
low energy electrons, which follow the ionization of the atom during the
LSP-nucleus collision. This possibility may be realized with the
technology of gaseous TPC detectors \cite{GV03}. In fact the WIMP-nucleus scattering
leads: (i) The nuclear recoil and (ii) to nuclear recoil with atomic excitation. So far most
CDM searches have been made by the inclusive processes, (i) and (ii), employing solid
detectors. We propose that the produced electrons in atomic excitation (ii) should be 
studied by exclusive measurements. 

\section{The effect of bound electrons}
\label{bounde}
 The differential cross section for the LSP nucleus scattering leading
  to the emission of electrons in the case of non relativistic neutralinos
   takes the form:
\barr
 d\sigma({\bf k})&=&\frac{1}{\upsilon}\frac{m_e}{E_e}|\textsl{M}|^2
             \frac{d^3{\bf q}}{(2\pi)^3}\frac{d^3{\bf k}}{(2\pi)^3}
             (2\pi)^3 \frac{1}{2(2\ell+1)} \\
 \nonumber
\sum_{n\ell m} p_{n\ell} &&[\tilde{\phi}_{n\ell m}({\bf k})]^2
             2\pi \delta(T_{\chi}+\epsilon_{n\ell}-T-\frac{q^2}{2m_A}-\frac{({\bf
             p} _{\chi}-{\bf k}-{\bf q})^2}{2m_{\chi}}),
\label{cross1}
  \earr
   where $\upsilon$,  $T_x$ and ${\bf p}_{\chi}$ are the oncoming LSP
   velocity, energy and momentum respectively, while ${\bf q}$ is the momentum
   transferred to the nucleus. $\textsl{M}$ is the invariant
amplitude, known from the standard neutralino nucleus cross
section, $T$ and ${\bf k}$ are the kinetic energy and momentum of
the outgoing electron
and $\epsilon_{n\ell}$ is the binding energy of the initial
electron. $\tilde{\phi}_{n\ell m}({\bf k})$ is the Fourier
transform of the bound electron wave function, i.e its wave
function in momentum space. $p_{n\ell}$ is the probability of
finding the electron in the  ${n,\ell}$ orbit.

In the expression above and in what follows our normalization will
consist of one electron per atom, to be compared with the cross
section per nucleus of the standard experiments.

The binding energies for hydrogenic wave functions are given by:
\beq \epsilon_n= - \frac{m_e (Z\alpha)^2}{2 n^2} ~~.
\label{Balmer} \eeq
 One, of course, may have to introduce effective charges, especially for heavy
 atoms. Some hydrogenic wave functions for light systems
 are going to be given below.

 After summing over the m-substates we find that:
 \beq
 \frac{1}{2(2\ell+1)}\sum_{m}|\tilde{\phi}_{n\ell m}({\bf
 k})|^2=\frac{1}{8 \pi}|\tilde{\phi}_{n\ell}(k)|^2.
 \label{Rkl}
 \eeq
 Due to the delta function the integration over the momentum
  q is trivial and yields
  \beq
  \int q^2dq \delta(T_{\chi}+\epsilon_{n\ell}-T-\frac{q^2}{2m_A}-\frac{({\bf
             p}_{\chi}-{\bf k}- {\bf q})^2}{2m_\chi})=
             \upsilon K \mu^2_r \frac{(\xi+\Lambda)^2}
             {\Lambda},
\label{mom1}
 \eeq
 with   $\mu_r$  the LSP-nucleus reduced mass and
 \beq
  \Lambda= \sqrt{ \xi^2+\frac{m_{\chi}}{\mu_r}(\frac{1}{K^2}-1)-
             \frac{m_{\chi}}{\mu_r K^2}\frac{(T-\epsilon_{n\ell})}{T_{\chi}}},
  \label {mom2}
  \eeq
with
\beq
 {\bf K}=\frac{{\bf p}_{\chi}-{\bf k}}{p_{\chi}},
K=\frac{\sqrt{p_{\chi}^2+k^2-2k p_{\chi} \xi_1}}{p_{\chi}},
\xi_1=\hat{p}_{\chi} \cdot  \hat{k},\xi=\hat{q} \cdot  \hat{K}
\label{K}
\eeq
 In order to avoid any complications arising from questions regarding the
 allowed SUSY parameter space, a subject extensively studied anyway
  \cite{Jung}-\cite{WELLS},
   we will present our results normalized to
  the standard neutralino nucleus cross section.
   The thus obtained branching  ratios are
   independent of all parameters of supersymmetry except the neutralino
   mass. The numerical results given here apply in the case of the coherent
   mode. If, however, we limit ourselves to the ratios of the
   relevant cross sections,
    we do not expect  substantial changes in the case of the
  spin induced process.

  With these ingredients we find that the ratio of the cross
 section with ionization divided by that of the standard
 neutralino-nucleus elastic scattering, nuclear recoil experiments (nrec), takes the form:
 \barr
 \frac {d\sigma(T)}{\sigma_{nrec}}&=&\frac{1}{4}\sum_{n\ell}p_{n\ell}
|\tilde{\phi_{n\ell}}(2m_eT)|^2
  \nonumber\\
  && \frac{\int _{-1} ^1
d\xi_1\int_{\xi_L} ^1 d \xi K \frac{(\xi+\Lambda)^2}{\Lambda}
 [F(\mu_r\upsilon (\xi+\Lambda))]^2}{\int_0^1 2 \xi d \xi
 [F(2 \mu_r \upsilon \xi)]^2}{m_e}k  dT,
 \label{ratio1}
  \earr
where $\xi_L=
\sqrt{\frac{m_{\chi}}{\mu_r}[1+\frac{1}{K^2}(\frac{T-\epsilon_{n\ell}}{T_{\chi}}-1)]}$,
$2\frac{\mu_r}{m_{\chi}} p_{\chi}\xi=2\mu_r\upsilon\xi$ is
 the momentum $q$ transferred to the nucleus and
 $F(q)$ is the nuclear form factor. The outgoing electron energy
lies in the range $0\leq T\leq
\frac{\mu_r}{m_{\chi}}T_{\chi}-\epsilon_{n\ell}$. The expression for the 
standard cross section $\sigma_{nrec}$ will be given below (Eq. \ref{slsp}).

Since the momentum of the outgoing electron is much smaller than
the momentum of the oncoming neutralino, i.e. $K\approx 1$, the
integration over $\xi_1$ can be trivially performed. Furthermore,
if the effect of the nuclear form factor can be neglected, the
integration over $\xi$ can be performed analytically. Thus we get:
\barr
 \frac {d\sigma(T)}{\sigma_{nrec}}&=&\frac{1}{2} \sum_{n\ell}p_{n\ell}
|\tilde{\phi_{n\ell}}(\sqrt{2m_eT})|^2 
[1-(\frac{m_{\chi}}{\mu_r}\frac{(T-\epsilon_{n\ell})}{T_{\chi}})
\nonumber\\&+&
 \sqrt{1-(\frac{m_{\chi}}{\mu_r}
 \frac{(T-\epsilon_{n\ell})}{T_{\chi}})}]
                               {m_e}\sqrt{2m_eT} dT .
 \label{ratio2}
  \earr
Otherwise the angular integrations can only be done numerically.
Furthermore integrating numerically the above expression over the
electron spectrum we obtain the total cross section.
 One, of course,  must convolute the above expression with the
 velocity distribution (see next section) to obtain both the differential
 rate as well the total rate as a function of the
 neutralino mass.

Before proceeding further we will give an estimate of the expected
standard elastic scattering recoil coherent events. The LSP- nucleus cross-section
takes the form:
  \beq
\sigma_{nrec}=(\frac{\mu_r}{\mu_r (p)})^2 A^2 \sigma_p
 \int_0^1 2 d \xi [F(2\mu_r \upsilon \xi)]^2,
 \label{slsp}
  \eeq
where $\sigma_p$ is the LSP-proton cross section and  $\mu_r
(p)\approx m_p$ is the reduced mass for the proton-LSP
system. From this expression we see the coherent effect of all nucleons arising
from the elementary scalar interaction. In the case of a light target like $^{20}$Ne the effect of
the form factor is not large. The event rate, corresponding to an
LSP velocity $ \sqrt{\left<\upsilon^2 \right>}$, takes the form:
\begin{equation}
R =\frac{dN}{dt} =\frac{\rho (0)}{m_{\chi}}
 \frac{m}{A m_p} \sqrt{\left<\upsilon^2 \right>} (\frac{\mu_r}{\mu_r (p)})^2 A^2 \sigma_p ,
 \label{2.17}
\end{equation}
 where   $\rho (0) = 0.3 GeV/cm^3$ is the LSP density in our
 vicinity and m is the detector mass. The upper limit of the current experiments is
 $\sigma_p\leq10^{-5}pb$. Using this limit with $m_{\chi}=100GeV$
 and  $\sqrt{\left<\upsilon^2 \right>}=270~km/s$,
 we get for the A=20 system
 $$\frac{dN}{dt}\approx 8~ events/(y.kg-target).$$
 This rate will be become $\approx 10~ events/(y.kg-target)$, if the folding
 with the LSP velocity distribution is taken into account (see sec. \ref{results}).
\section{ A brief discussion of the backgrounds}
 We must begin by emphasizing that the standard dark matter experiments have thus far consisted 
of inclusive measurements employing solid detectors, where the electrons and the 
recoil nuclei have not been separated. These measure continuum energy spectra of
the recoiling nuclei with an energy resolution, which is not very good. Most experiments
are employing neutrons with energy in the range of $5-10$ MeV with a resolution of a few
$0.1$ MeV, which is much larger than the electron energy. Then the electron contribution is
hidden. The present paper is concerned
with an exclusive measurement, whereby the electron signal is separated from the nuclear
recoil signal. The electron rate is about $10\%$ of the nuclear recoil rate. The actual
nuclear recoil rate is reduced by about $30-50~\%$ depending the detector ienergy cutoff.
 It thus 
appears that the electron rate could be as high as $1/3$ of the nuclear recoil rate. This makes
it a realistic proposal, if the BG problems are under control. The nice feature of an
 exclusive experiment is that one can measure simultaneously the electrons and the nuclear
recoils, substantially reducing this way the BG rate. 

 To circumvent the BG problems one must use high purity materials and run the experiment
underground. The main source of background is neutrons and this applies both in the
 standard experiments
as well as the present novel proposal.
It thus appears that background events of primary concern to the present work
 can arise  from two main sources:
 \begin{enumerate}
 \item Reaction induced Auger electrons. These can arise from the decay of the
 inner electron holes in the daughter atom. The decay width has been estimated
to be $1.2\times 10^{-9}$keV for $^{20}$Ne. These electrons can have energies in the
 region of interest. In
  the case of $^{20}$Ne they are expected at energies of $0.77,~0.80$
and $0.83~keV$ depending on which two electrons fill this hole and
produce the Auger electron \cite{GV03}. The reaction induced 1s
holes, however, occur with a small probability, see Fig.
\ref{dnof-f}
One may also have background electrons, produced from impurities
which may lead to betas and photoelectrons producing $\gamma$ rays.
Anyway it has been estimated that such a background is under
control \cite{GV03}.
\item Radioactive Impurities (RI). RI 's give rise to  $\beta,\gamma,\alpha$ BG's.
 As it is well known,
recent recoil detectors can separate the recoil signal from electron backgrounds of RI
$\beta,\gamma$'s. Similarly the electron tracking detectors select electron signals. It is
quite realistic, as shown both in the recent solar $\nu$ detectors and the planned $\beta \beta$ 
experiments, to build detectors at the level of ppt of U-Th impurities (ppt$=10^{-12}$). Note
that most RI BG's deposit $\beta,\gamma$ energies at multi-sites. Then the BG rate for the
 electron detectors with impurities at the level of a few ppt is estimated at $30$ per year
 per ton, by a stringent selection 
of DM single-site signals from the RI multi-site events. This is orders of magnitude less than
 the present limit ($8$/year/kg) of the WIMP's. It should also be noted that this limit is 2-3
 orders of magnitude less than the electron signal rates ($1\sim 10$ per year per ton) and the
 electron BG rates of the planned electron detectors for real time solar p-p $\nu$'s and
 $\beta \beta$  experiments. In any case it seems feasible to build detectors with BG rates well
below the the signal rates with $10^{-5}$pb. Furthermore since the electron energy signal is always
 accompanied by the recoil signal, the coincidence measurement of both the electron tracks and
and the recoil signals will make this experiment almost free from BG events.
\item  Electron backgrounds due to RI. These are under control \cite{GV03}. Most electrons from RI's are
$\beta$ rays followed by $\gamma$  rays and Compton electrons followed by Compton $\gamma$ rays.
These deposit energy at multi-sites and, thus, they can be  eliminated by selecting the single-site
 event from
 WIMP's. With a purity level of the order of $\mu$Bq per Kg one gets a BG event rate of less than
 $0.1$ per year per kg , after a stringent signal selection. This is far below the present WIMP signal
 rate ($8$ per y per kg). 
\item Neutron induced events. This source is also present in the
standard nuclear recoil experiments and it should be dealt with in both.
Such neutron events can be
deciphered by studying the reaction products. We will, however,
estimate the event rate due to neutrons using the distribution for
muon produced neutrons given at 2000 m w.e. by Gaitskell
\cite{GAITSKELL}.
\begin{itemize}
\item Neutron electron scattering due to the neutron magnetic
dipole moment.

 This process is very special in the sense that the electron interacts
  directly with the neutron. Thus the magnetic moment interaction does not suffer from
  suppression at high electron energies due to the bound electron form
  factor,
  see Figs \ref{dnecross}- \ref{dnerate} obtained as discussed in the
   appendix.
\item  Electron ionization due to neutron-nucleus elastic
scattering.

 This process is characterized by kinematics similar to the
 reaction we are interested in. For the model discussed in the appendix
 involving a Yukawa interaction, see Fig. \ref{yukpot}, we obtain the results
 shown in Figs \ref{yuknn1}-\ref{yuknn2}.
 \end{itemize}
  It appears that both of these backgrounds yield rates that could be troublesome.
 In practice the rate gets smaller by 2-3 orders of magnitude at typical underground
  labs with 4000 - 6000 m w.e.
  Active neutron shields for incoming and outgoing neutrons reduce further the BG rate.
  Accordingly the BG rate due to the neutrons can be negligibly small in realistic cases.
  These BG problems are further discussed in the Appendix.
   \end{enumerate}

\section{Some  results}
\label{results}
   We will now apply the above formalism for a typical case, namely
  $m_{\chi}=100~GeV$. We will employ a Maxwell-Boltzmann LSP  velocity distribution
  with respect to the galactic center \cite{JDV03}, namely:
  \beq
  f({\bf v})=\frac{1}{(v_0 \sqrt{\pi})^3}e^{-(v^2/v_0^2)},
  \label{mbgc}
  \eeq
  with $v_0=220~km/s$. On this we impose by hand an upper velocity
  bound (escape velocity), $v_{esc}=2.84v_0$. We then transform this distribution
   to the lab
  frame, ${\bf v} \rightarrow {\bf v}+{\bf v}_0$, where ${\bf v}_0$ is the
   velocity of the sun around the center of the galaxy. We will not be concerned
   here with the motion of the the Earth around the sun, i.e. the modulation
   effect \cite{JDV03}.

  Folding both the numerator and the denominator of Eq. (\ref{ratio1}) with the LSP velocity
  distribution, after multiplying each with the LSP flux
  $\frac{\rho (0)}{m_{\chi}}
 \frac{m}{A m_p}\upsilon$, we obtain the differential ratio
 $\frac{1}{R}\frac{dR_e}{dT}$, with $R_e$ the rate for ionization.
   If we then integrate the final expression with respect to the
 electron energy, we obtain the ratio $\frac{R_{e}}{R}$. By doing the same in the case of
 Eq. (\ref{ratio2}), we obtain the analogous ratios
 $\frac{1}{R_0}\frac{dR_{e0}}{dT}$ and $\frac{R_{e0}}{R_0}$,
 associated with the case without nuclear form factor.

  We will apply our results on a light target, e.g. $^{19}$F,
  which is considered in the standard nuclear recoil experiments.
  The obtained results are essentially identical to those for
  $^{20}$Ne, which is a popular gaseous TPC counter currently
  being considered for detection of low energy neutrinos produced
  in triton decay.
  Our approach can, of course,
   also be applied in the case of the heavier
  targets, employed in current dark matter searches, but the atomic physics
  involved is much more complicated

\begin{figure}
\rotatebox{90}{\hspace{0.0cm}
 $\frac{1}{R0} \frac{dR_{e0}}{ dT}\rightarrow keV^{-1}$}
\includegraphics[height=.13\textheight]{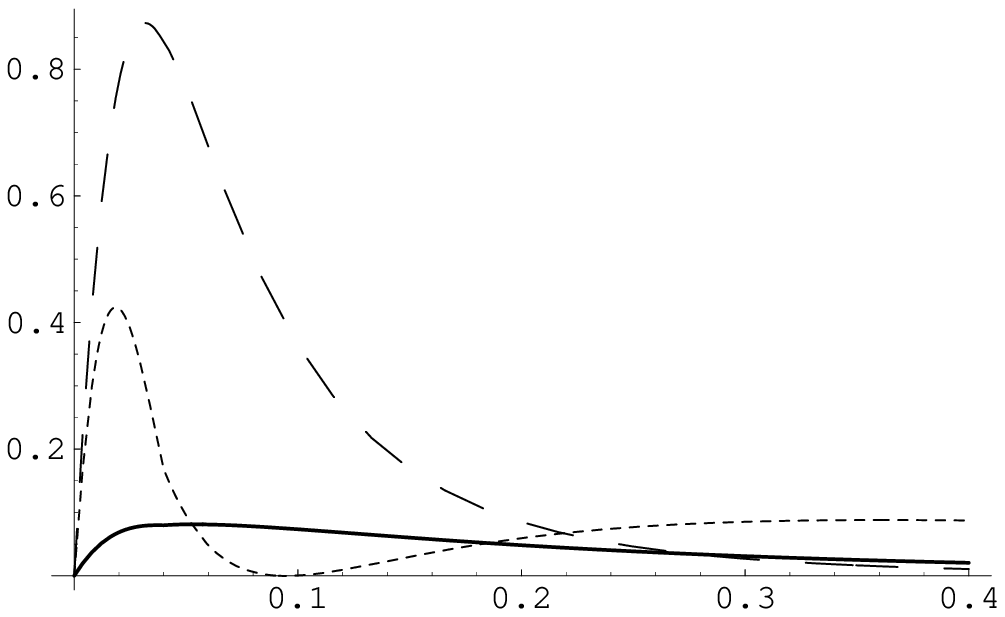}
{\hspace{0.0cm} {\tiny $T\rightarrow keV $}}
\rotatebox{90}{\hspace{0.0cm}
 $\frac{1}{R}\frac{dR_e}{ dT}\rightarrow keV^{-1}$}
\includegraphics[height=.13\textheight]{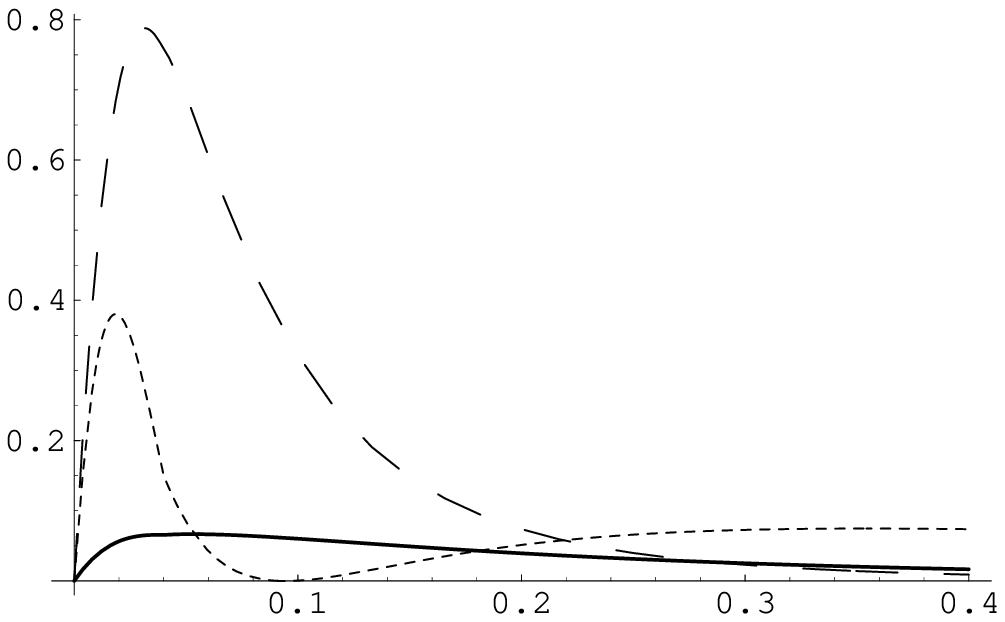}
{\hspace{0.0cm} {\tiny $T\rightarrow keV $}}
\caption{The differential rate, divided by the total rate
associated with the nuclear recoils, as a function of the electron
energy T (in $keV$). Each atomic orbit involved in the target
$^{20}$Ne is included separately.
  The results exhibited are obtained
without nuclear form factor on the left and with nuclear form
factor on the right. The full line, the short-dashed line and the
long-dashed line correspond to the orbits $1s~,~2s~$ and $2p$
respectively.
 \label{dnof-f}}
\end{figure}

\begin{figure}
\rotatebox{90}{\hspace{0.0cm}
 $\frac{1}{R0}\frac{dR_{e0}}{ dT}\rightarrow keV^{-1}$}
\includegraphics[height=.13\textheight]{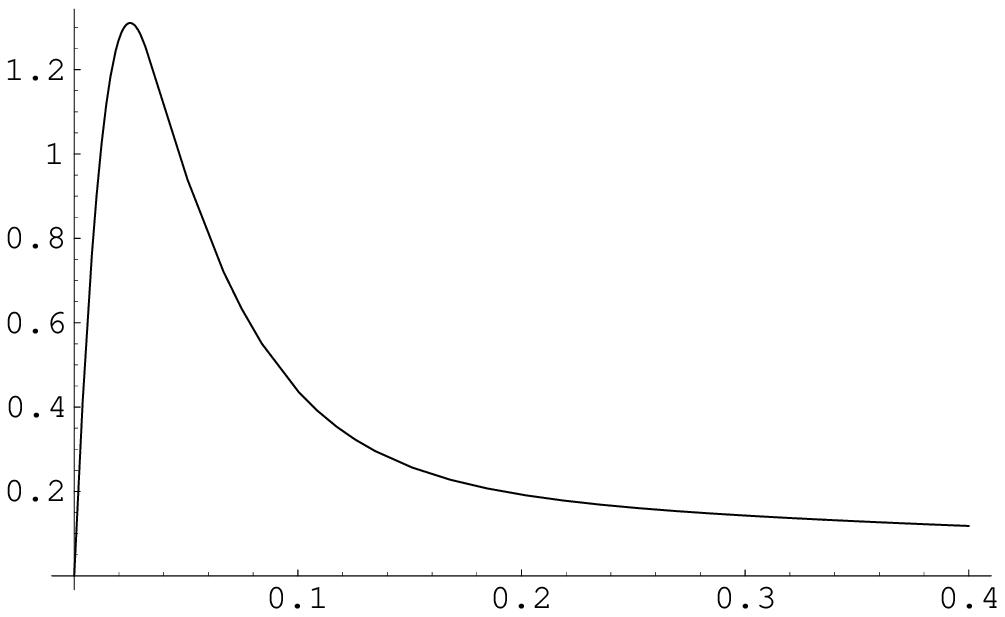}
{\hspace{0.0cm} {\tiny $T\rightarrow keV $}}
\rotatebox{90}{\hspace{0.0cm}
 $\frac{1}{R}\frac{dR_e}{ dT}\rightarrow keV^{-1}$}
\includegraphics[height=.13\textheight]{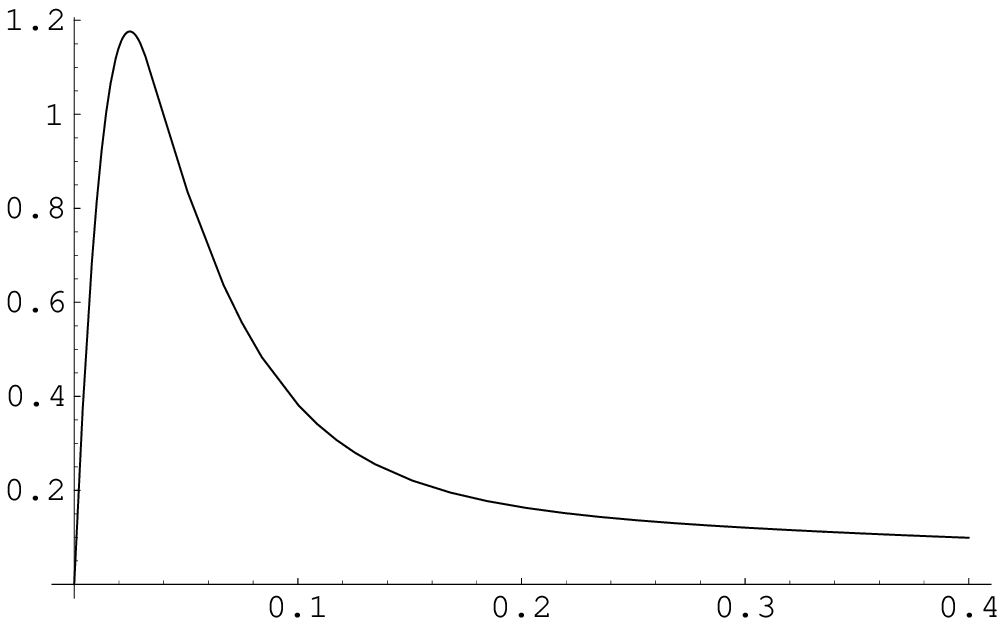}
{\hspace{0.0cm} {\tiny $T\rightarrow keV $}}
\caption{The same as in Fig. \ref{dnof-f} by summing the
contribution of all orbitals together.
 \label{nof-f}}
\end{figure}

\begin{figure}
\rotatebox{90}{\hspace{0.0cm}
 $\frac{R_{e0}}{R_0}\rightarrow$}
\includegraphics[height=.12\textheight]{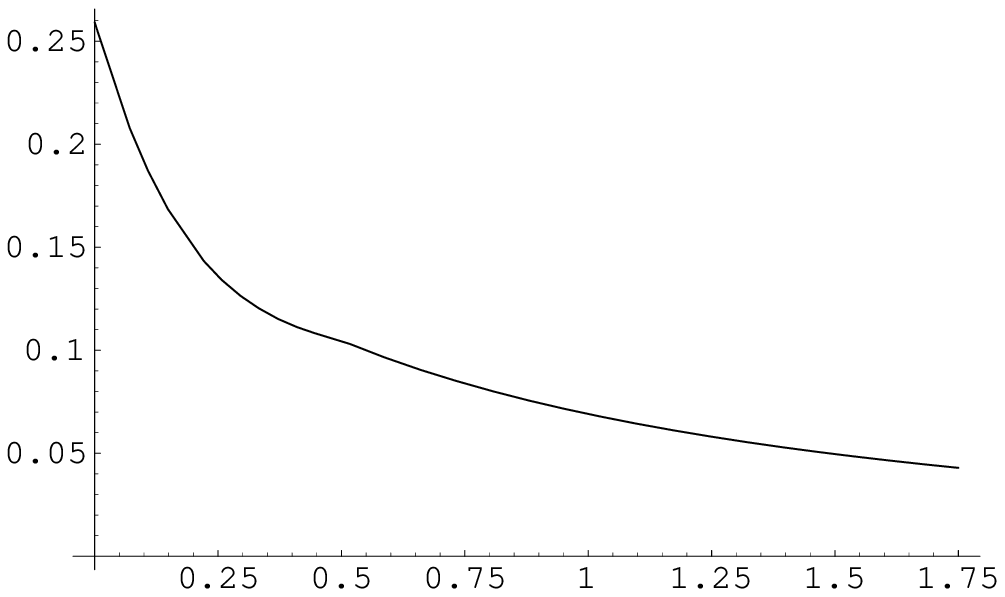}
{\hspace{0.0cm}{\tiny $E_{th} \rightarrow keV$}}
\rotatebox{90}{\hspace{0.0cm}
 $\frac{R_e}{R}\rightarrow$}
\includegraphics[height=.12\textheight]{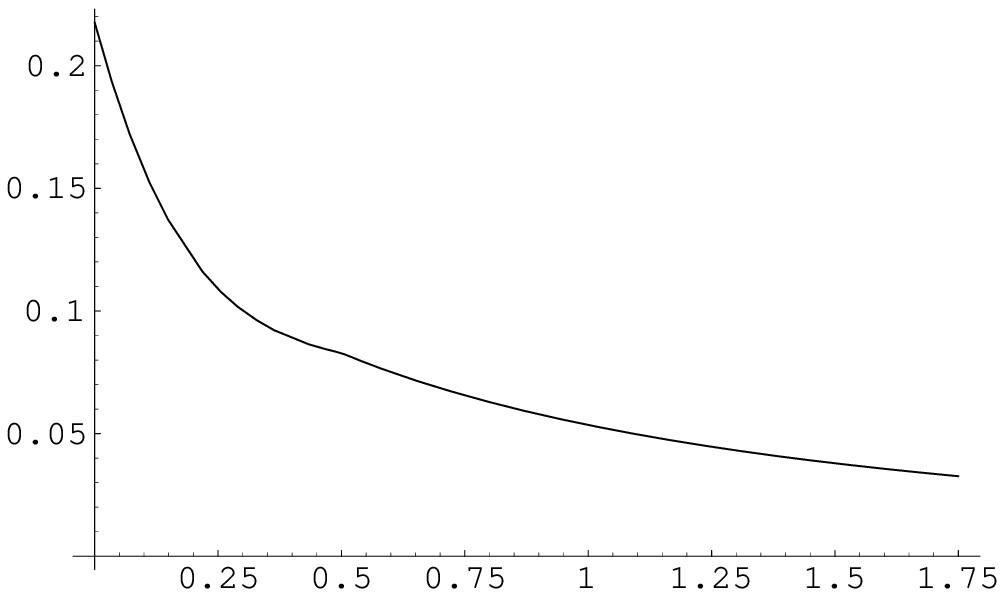}
{\hspace{0.0cm} {\tiny $E_{th}\rightarrow keV $}}
\caption{The total rate for ionization divided by the standard
nuclear recoil rate as a function of the threshold energy. No low
energy cutoff imposed by the detector was included in the
computation of the standard rate. The estimated rate for the
proposed mechanism is obtained by multiplying this branching ratio
with the standard event rate. In the case of $^{20}$Ne this about
10 events/(kg-y), obtained from  the upper bound of the cross
section as derived from the current experiments.
 \label{cnof-f}}
\end{figure}

 The radial functions $\tilde{\phi_{n\ell}}(k)$ will be given in terms of a
dimensionless variable $b=\frac{nka_{B}}{2z}$, by writing
$\tilde{\phi_{n\ell}}(k)\Rightarrow R_{n\ell}(b)$, with
$R_{n\ell}^2(b)$  given by:
 \beq
  R^2_{1s}=\frac{256}{{\left( 1 + 4\,b^2
\right) }^4\,\pi },
\label{rad1} \eeq \beq R^2_{2s}=\frac{512\,{\left( 1 - 4\,b^2
\right) }^2}{{\left( 1 + 4\,b^2 \right) }^4\,\pi } ~,~
 R^2_{2p}=
\frac{16384b^2}{3\,{\left( 1 + 4\,b^2 \right) }^6\,\pi}.
\label{rad3}
 \eeq
 The binding energies employed are given in the above order by
  \cite{GOUNARIS}:
\beq \epsilon_{1s}=-0.870~
,~\epsilon_{2s}=-0.048~,~\epsilon_{2p}=-0.021 .
 \label{bener}
 \eeq
 In Fig.\ref{dnof-f} we show the differential rate of our process,
 divided by the total  nuclear recoil event rate,
 for each orbit
 (see Eq. \ref{ratio2})
   without the inclusion of the nuclear form factor on the left
 and with  appropriate form
 factor \cite{DIVA00} on the right. In Fig. \ref{nof-f} we show the
 same quantity taking all orbitals together. These results were obtained
 considering one electron per atom and weighing
  each orbit with the probabilities $p_{n\ell}$:
  \beq
  p_{n\ell}=(2/10,2/10,6/10).
 \label{pnl}
 \eeq
     The branching ratio for the total rate leading to our process,
      obtained as discussed above,
  are shown in Fig. \ref{cnof-f}, as a function of the low energy
 detector cut-off (threshold). From this plot we see that the introduction of the
 nuclear form factor, which we view as quite reliable
 \cite{DIVA00}, does not substantially alter the branching ratio.
 This trend for the branching ratio may persist even for heavier
 nuclei, even though the individual rates may depend on the
 assumed form factor.

 From these plots we see that, even though the differential rate peaks at low energies,
there remains
 substantial strength above the electron energy of $0.2~keV$, which is the threshold energy 
used in
  MICROMEGAS  detectors, like the one recently
 \cite{GV03} proposed.
\section{Conclusions}

    Based on our results summarized in the previous section one can be optimistic
   about using the emitted electrons in the neutralino nucleus
   collisions for the direct detection of the LSP.  This novel process
   may be exploited by the planned TPC low energy electron
   detectors. By achieving   low energy thresholds of about $0.25~keV$, the
   branching ratios are approximately 10 percent. They can be even
   larger,
   if one includes low energy cutoffs imposed by the detectors in the
   standard experiments, not included in the above estimate.

    As we have seen the background problems associated with the proposed
    mechanism are not worse than those entering the standard experiments. In
    any case coincidence experiments with
   x-rays, produced following the de-excitation of the residual atom, may help reduce
   the background events to extremely low levels.
\begin{figure}
\rotatebox{90}{\hspace{0.0cm} $\frac{d\sigma_{ne}}{dT}
\rightarrow~b/keV$}
\includegraphics[height=.12\textheight]{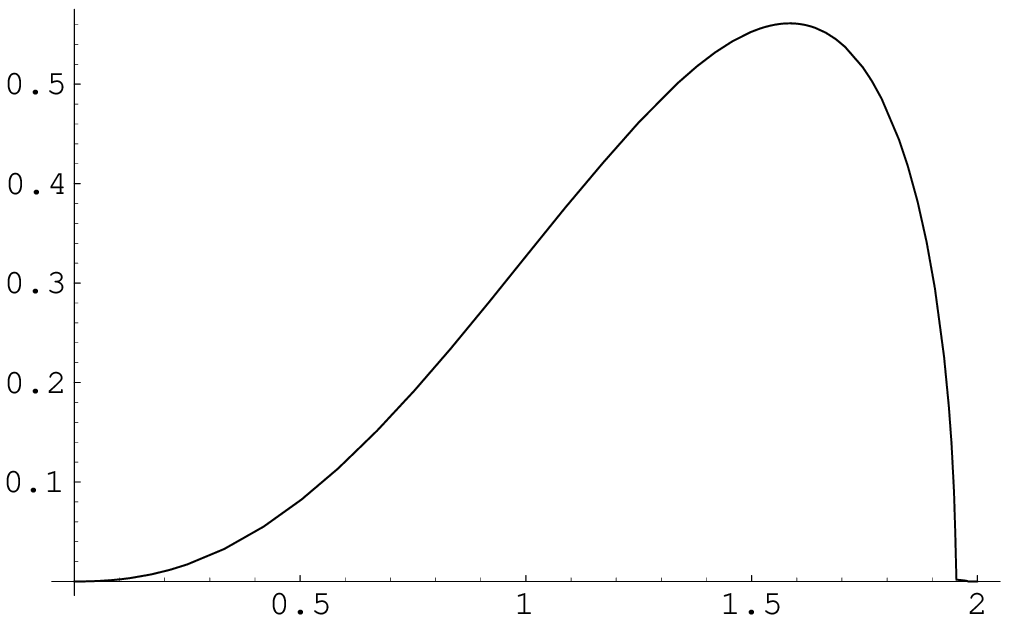}
{\hspace{0.0cm}{\small $T \rightarrow keV$}}
\rotatebox{90}{\hspace{0.0cm} $\frac{d\sigma_{ne}}{d\epsilon_n}
\rightarrow~b/keV$}
\includegraphics[height=.12\textheight]{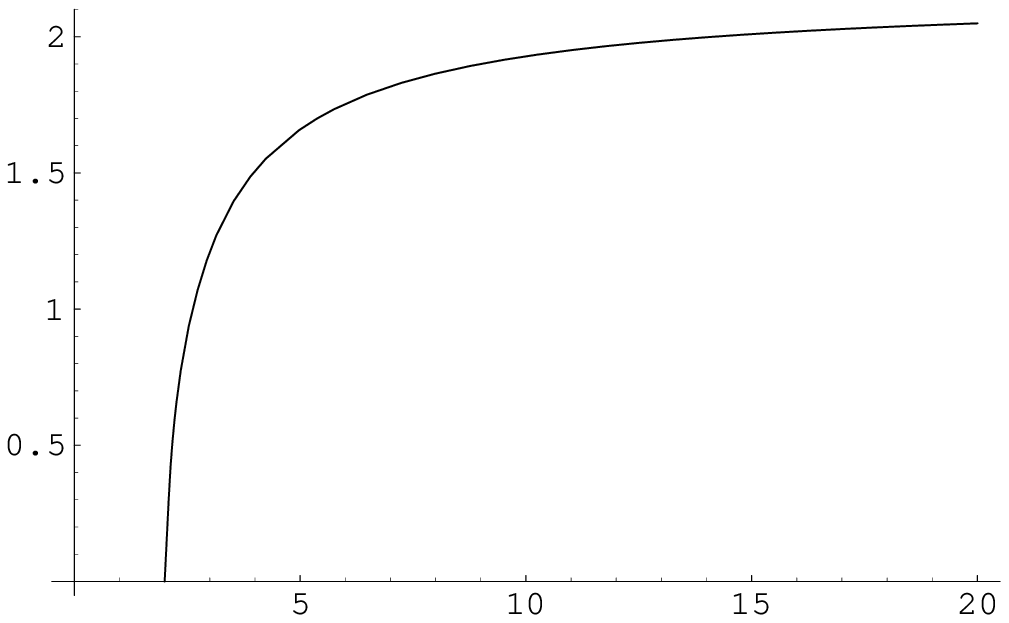}
 {\hspace{0.0cm}{\small $\epsilon_n \rightarrow keV$}}
\caption{The differential cross section for neutron electron
scattering due to the neutron magnetic moment in the case
$^{20}$Ne. The results on the left correspond to an incident
neutron energy of $2 ~keV$ plotted as a function of the outgoing
electron energy for electrons bound in the $1s,~2s$ and $2p$
orbitals. The results on the right correspond to an outgoing
electron of $2~ keV $as a function of the incident neutron energy.
 \label{dnecross}}
\end{figure}
\begin{figure}
\rotatebox{90}{\hspace{0.0cm} $\frac{dR_{ne}}{dT}
\rightarrow~$events/(kg.y.keV}
\includegraphics[height=.2\textheight]{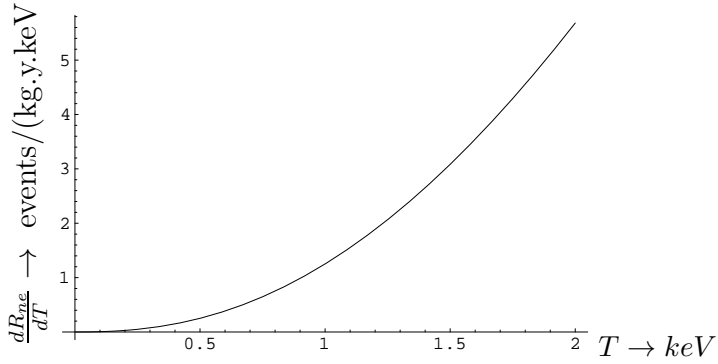}
{\hspace{0.0cm}{\small $T \rightarrow keV$}} \caption{The
differential event rate for electron ionization in neutron
electron scattering, which is due to the neutron magnetic moment,
in the case $^{20}Ne$ as a function of the outgoing electron
energy. Electrons bound in the $1s,~2s$ and $2p$ orbitals were
considered . The results presented were obtained using
distribution of Gaitskell for muon produced neutrons. All neutrons
were included, but the event rate reached saturation in about 500
MeV.
 \label{dnerate}}
\end{figure}
\begin{figure}
\rotatebox{90}{\hspace{0.0cm} Potential $\rightarrow \alpha_s A
m_{\pi}^{-1}$}
\includegraphics[height=.2\textheight]{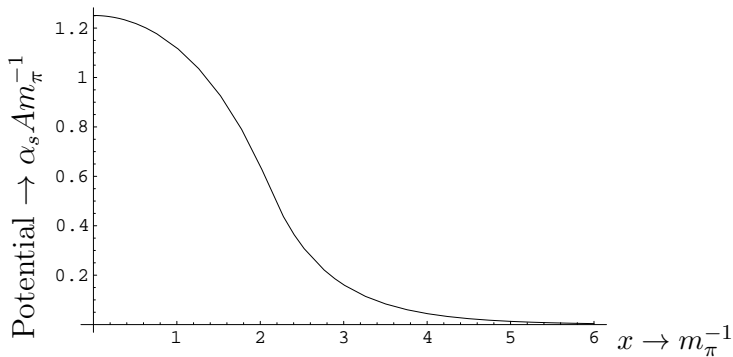}
 {\hspace{0.0cm}{\small $x\rightarrow m_{\pi}^{-1}$}}
\caption{The strong interaction (Yukawa like) potential felt by
the elastically scattered neutrons for the $A=20$ system.
 \label{yukpot}}
\end{figure}
\begin{figure}
\rotatebox{90}{\hspace{0.0cm}
$\frac{d\sigma_{N}}{dT}\rightarrow~(b/keV) \alpha_s^2$}
\includegraphics[height=.2\textheight]{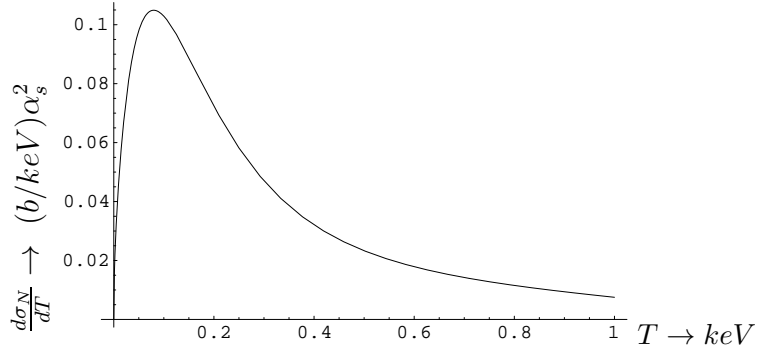}
{\hspace{0.0cm}{\small $T \rightarrow keV$}}
 \caption{
The differential cross section for neutron electron scattering due
to the strong interaction. The value $\alpha_s=1.0$ was used in
the numerical calculation. Otherwise the notation is the same as
in Fig. \ref{dnecross}.
 \label{yuknn1}}
\end{figure}
\begin{figure}
\rotatebox{90}{\hspace{0.0cm} $\frac{dR_{ne}}{\alpha^2_sdT}
\rightarrow~$events/(kg.y.keV}
 \includegraphics[height=.25\textheight]{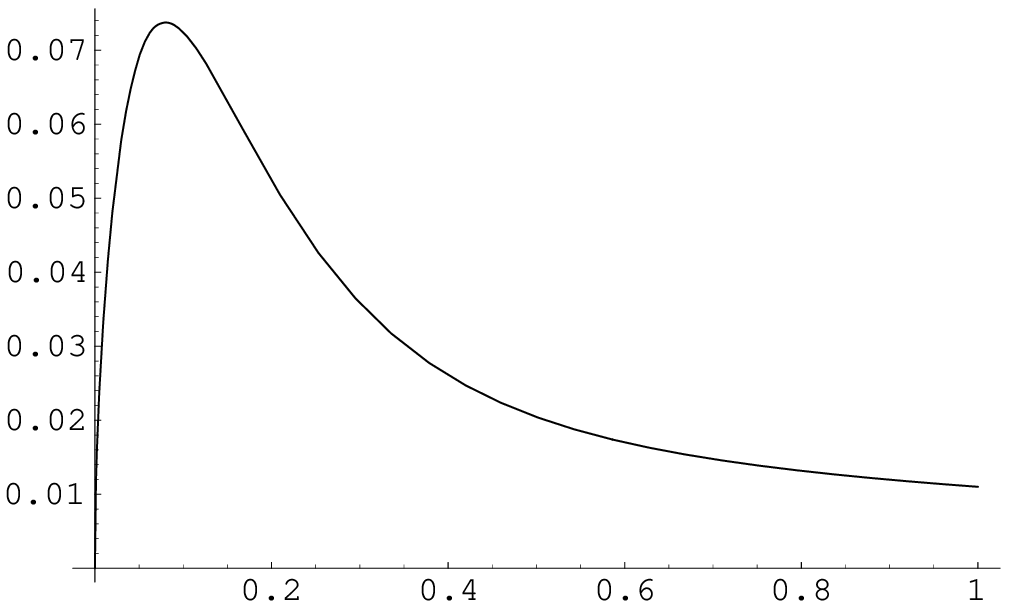}
{\hspace{0.0cm}{\small $T \rightarrow keV$}}
 \caption{
The differential rate for electron ionization in elastic neutron
nucleus scattering. Otherwise the notation is the same as in Fig.
\ref{dnerate}
 \label{yuknn2}}
\end{figure}

Acknowledgments: This work   was supported in part by the European
Union under the contracts RTN No HPRN-CT-2000-00148 and MRTN-CT-2004-503369.
  One of the authors (JDV) is indebted to
Professors K. Zioutas for a careful reading of the manuscript and
Y. Giomataris and N. Alexandropoulos for their useful comments
regarding experimental issues related to this work.

 \appendix{Appendix: Some further discussion on the  background.}
\renewcommand{\theequation}{A.\arabic{equation}}

 In this section we are going to  briefly discuss the possible background
 (BG) events. A more detailed discussion will be given elsewhere.
  RI impurities in all detector components should be as low as
 possible to avoid all kinds of BG electrons from $\beta$ and
 $\gamma$ rays from RI. In practice the Ur and Th impurities
in the rock at underground laboratories
 can be reduced to be less than $10^{-16}$ in weight leading to a BG
 event rate of 0.03 per Kg per year, much lower than the DM event
 rate of the present concern.

 Neutrons in the 10 MeV -1 GeV region are of the order of 1 per $m^2$
 per year at underground laboratories of $4000 m$ water
 equivalent. Lower energy neutrons arising from nuclear fission
 of Ur-Th isotopes around the detector can be slowed down and be
 absorbed.

 This source is also present in the
standard nuclear recoil experiments. Such neutron events can be
deciphered by studying the reaction products. We will, however,
estimate the event rate due to neutrons using the distribution for
muon produced neutrons given by Gaitskell \cite{GAITSKELL},
approximated by:.
 \beq
 f[e_n]=(1.72e^{-[e_n/(241MeV)]}+8.27e^{-[e_n/(55.5MeV)]})/MeV.y.m^2~.
 \label{fen} \eeq
 We distinguish the following cases:
\begin{itemize}
\item Neutron electron scattering due to the neutron magnetic
dipole moment.
The leading term in this interaction is of the form:
 \beq
 \Omega({\bf r})=\sqrt{4\pi \alpha} \frac{\mu_n}{2m_e}[\sigma_e \times \sigma_n]^2 
 \cdot \sqrt{4 \pi} Y^2 (\hat{r})\frac{1}{r^3} .
 \label{muemag}
  \eeq
  Thus one must evaluate the orbital matrix element
  \beq
  \int d^3{\bf x}\Omega(\sqrt{2}{\bf
  x})e^{i{\bf k}.{\bf x}\sqrt{2}}\int d^3{\bf y}\phi_b(\frac{{\bf x}-{\bf
  y}}{\sqrt{2}}),
   \label{intme}
    \eeq
 where $\phi_b$ is the bound electron wave function and ${\bf
 k}$ is the momentum of the outgoing electron. The evaluation of the
 above integral proceeds by a multipole
 expansion. For bound electrons in  s-states the only multipole
 appearing is the quadrupole, while for p-states one can have
 both a dipole and octapole term with the dipole dominating.
 One can then compute the spin matrix elements and calculate the
 differential cross section as a function of the outgoing electron
 energy. One should use a proper  neutron energy distribution function.
  For orientation purposes we
 calculated the differential cross section for a  neutron energy of
 $\epsilon_n=2~keV$, which leads to a maximum electron
 energy $\epsilon_n+\epsilon_{n\ell}$.
 The obtained results are shown in Fig. \ref{dnecross}. We
 see that the shape of the differential cross section is very different from
 that of the interesting electrons. Furthermore such events can be
 rejected by the subsequent detection of the neutron. The event
 rate, obtained by using the Gaitskell neutron distribution, is
 given in Fig. \ref{dnerate}.
 \item Neutron nucleus elastic scattering. This can lead to electron
 ejection in pretty much the same way as the LSP-Nucleus elastic
 scattering.
 Background events can be rejected by the detection of the scattere neutron.

 Proceeding in a manner analogous to that described in
 sec. \ref{bounde} we find:
  \barr
 d\sigma({\bf k})&=&\frac{1}{\upsilon}\frac{m_e}{E_e}|\textsl{M}|^2
             \frac{d^3{\bf p^{'}}_n}{(2\pi)^3}\frac{d^3{\bf k}}{(2\pi)^3}
             (2\pi)^3 \frac{1}{2(2\ell+1)} \\
 \nonumber
\sum_{n\ell m} p_{n\ell} &&[\tilde{\phi}_{n\ell m}({\bf k})]^2
             2\pi \delta(\epsilon_n+\epsilon_{n\ell}-T-\epsilon^{'}_n-\frac{({\bf
             p} _n-{\bf k}-{\bf p}_n^{'})^2}{2m_A}),
\label{ncross1}
  \earr
with the amplitude M being just the Fourier transform of the
potential generated by the A nucleons, which is felt by the
neutrons:
 \beq
 M=\int d^3{\bf r} e^{i({\bf p}^{'}_n-{\bf p}_n).{\bf r}}
 V_A(r)= {\tilde V_A}(|{\bf p}^{'}_n-{\bf p}_n|).
 \label{fourier1}
 \eeq
 Ignoring the energy taken away by the nucleus we finally find:
 \barr
  d\sigma_{(n,(e,n))}&&(\epsilon_n,T)=
   \frac{1}{\pi \sqrt{2}}
  m^2_n m_e \sqrt{m_eT} dT \\
  \nonumber &&\sum_{n\ell}p_{n\ell}
|\tilde{\phi_{n\ell}} (\sqrt{2m_eT})|^2
 F_A(\epsilon,\frac{\epsilon_{n\ell}-T}{\epsilon_n}),
 \label {fourier2}
  \earr

with $F_A(\epsilon_n,x)$ given by:
 \beq
 F_A(\epsilon_n,x)=\int_{1-\sqrt{1-x}}^{1+\sqrt{1-x}}tdt \left[ \tilde{V}_A\left(
 b\sqrt{2m_n\epsilon_n}~t \right) \right]^2,
 \label{intxi}
 \eeq
 where $b$ is a length parameter characterizing the range of the potential.

  The essential input in this case is the neutron-nucleus elastic
  scattering potential $V_A$, which can be determined
  phenomenologically.

   Proceeding further the elastic neutron
  nucleus cross-section can be obtained in an analogous fashion:
 \beq
 d\sigma_{(n,n)}(\epsilon_n)= \frac{1}{2 \pi}\frac{
 m_n}{2\epsilon_n}\int_0^{\epsilon_n}(\epsilon_n-x)dx \left[ \tilde{V}_A\left(
 b(\epsilon_n-x) \right) \right]^2.
 \label {fourier6}
  \eeq

 In other words the evaluation of the interesting cross-section  requires a
 complicated analysis of the elastic neutron nucleus
 cross-section. Thus
  for the purposes of the present work we have decided to obtain
  the potential by folding the
  Yukawa interaction with a range $b=m_{\pi}^{-1}$ with a constant
   nuclear density . The thus obtained
  potential is shown in \ref{yukpot}. The obtained elastic neutron nucleus
  rate is presented in Fig. \ref{yuknn2}. We
  see that this background contribution is peaked at very low electron energy.
  \item Compound nucleus formation by neutron capture. Such
  background events can be rejected by detecting the decay
  $\gamma$-rays  of the compound nucleus.

 In this case, if one ignores the tiny energy taken away by the outgoing
 nucleus, for a neutron of energy
 $\epsilon_n$ the electron energy is fixed, $T=\epsilon_n+\epsilon_{n\ell}+\Delta-E_x$,
 where $\Delta$ is the available energy and $E_x$ is the energy of the
 of the compound nuclear state. For thermal neutrons the electron
 energy is essentially fixed by the energy of the compound nucleus
 $T \approx \epsilon_{n\ell}+\Delta-E_x$. Thus this source does
  not pose a serious problem.

 Anyway, if one wants, one can cast the relevant  cross section in the form:
 \barr
 \sigma(\epsilon_n)&=&
 \frac{m_e}{2(2\pi)^2}\sqrt{\frac{m_n}{2\epsilon_n}}
   \sum_{n\ell,x}p_{n\ell}
   \sqrt{2m_e(\epsilon_n+\epsilon_{n\ell}+\Delta-E_x)}\\
   \nonumber
   &&|\tilde{\phi_{n\ell}} (\sqrt{2m_e(\epsilon_n+\epsilon_{n\ell}+\Delta-E_x)})|^2
   |M(p_n,x)|^2,
  \label {compound1}
  \earr
 where $M(p_n,x)$ is the amplitude associated with the standard
 compound nucleus formation. Noting that the captured neutrons are
 thermal one finds that the event rate for this process can be related to the
 rate for neutron capture as follows:
\barr
 \frac{dN}{dt}&&\approx
   \frac{1}{2(2\pi)^3}\int_0^{\epsilon}d\epsilon_n m_e
   \sum_{n\ell,x}p_{n\ell}
  \sqrt{2m_e(\epsilon_n+\epsilon_{n\ell}+\Delta-E_x)}\\
  \nonumber
  && |\tilde{\phi_{n\ell}} (\sqrt{2m_e(\epsilon_n+\epsilon_{n\ell}+\Delta-E_x)})|^2
   \left(\frac{dN}{dt}\right)_{compound},
  \label {compound2}
  \earr
 where the last term in the above equation gives the rate for
 the compound nucleus formation and $\epsilon$ is the highest
 energy at which neutron capture can occur. We are not, however, going to
 elaborate further on this point.

   \end{itemize}

 The background rates have so far been evaluated by using the
 neutron flux at 2000 m w.e. given by Gaitskell. The BG rates are
 of the same order of magnitude as the true DM rates or smaller
 and are well reduced by requiring anti-coincidence with the
 reaction products. Furthermore present underground laboratories are typically
 at around 3000 m w.e. or deeper. Thus the neutron flux and the
 background rates at these depths get smaller by orders of
 magnitude compared to those at 2000 m w.e. It thus appears that the BG rates from
 neutrons are not serious.

 \end{document}